\documentclass[floatfix,showpacs,preprint,preprintnumbers,superscriptaddress,%
amsmath,amssymb,prc]{revtex4-1}

\usepackage{amsmath, amssymb}
\usepackage{graphicx}
\usepackage{times}
\usepackage{dcolumn}

\begin{document}

\preprint{}

\title{\boldmath Broken Time Reversal Symmetry in Superconducting Pr$_{1-x}$Ce$_{x}$Pt$_4$Ge$_{12}$}

\author{Jian~Zhang}
\affiliation{State Key Laboratory of Surface Physics, Department of Physics, Fudan University, Shanghai 200433, China}
\author{D.~E.~MacLaughlin}
\affiliation{Department of Physics and Astronomy, University of California, Riverside, California 92521, USA}
\author{A.~D.~Hillier}
\affiliation{ISIS facility, STFC Rutherford Appleton Laboratory, Harwell Science and Innovation Campus,\\ Chilton, Didcot, Oxon.\ OX11 0QX, United Kingdom}
\author{Z.~F.~Ding}
\affiliation{State Key Laboratory of Surface Physics, Department of Physics, Fudan University, Shanghai 200433, China}
\author{K.~Huang}
\affiliation{Department of Physics, University of California, San Diego, La Jolla, California 92093,USA}
\affiliation{State Key Laboratory of Surface Physics, Department of Physics, Fudan University, Shanghai 200433, China}
\author{M.~B.~Maple}
\affiliation{Department of Physics, University of California, San Diego, La Jolla, California 92093,USA}
\author{Lei~Shu}\thanks{leishu@fudan.edu.cn}
\affiliation{State Key Laboratory of Surface Physics, Department of Physics, Fudan University, Shanghai 200433, China}
\affiliation{Collaborative Innovation Center of Advanced Microstructures, Fudan University, Shanghai 200433, China}

\date{\today}

\begin{abstract}
We report results of zero-field muon spin relaxation experiments on the filled-skutterudite superconductors~Pr$_{1-x}$Ce$_{x}$Pt$_4$Ge$_{12}$, $x = 0$, 0.07, 0.1, and 0.2, to investigate the effect of Ce doping on broken time-reversal symmetry (TRS) in the superconducting state. In these alloys broken TRS is signaled by the onset of a spontaneous static local magnetic field~$B_s$ below the superconducting transition temperature. We find that $B_s$ decreases linearly with $x$ and $\to 0$ at $x \approx 0.4$, close to the concentration above which superconductivity is no longer observed. The (Pr,Ce)Pt$_4$Ge$_{12}$ and isostructural (Pr,La)Os$_4$Sb$_{12}$ alloy series both exhibit superconductivity with broken TRS, and in both the decrease of $B_s$ is proportional to the decrease of Pr concentration. This suggests that Pr-Pr intersite interactions are responsible for the broken TRS\@. The two alloy series differ in that the La-doped alloys are superconducting for all La concentrations, suggesting that in (Pr,Ce)Pt$_4$Ge$_{12}$ pair-breaking by Ce doping suppresses superconductivity. For all $x$ the dynamic muon spin relaxation rate decreases somewhat in the superconducting state. This may be due to Korringa relaxation by conduction electrons, which is reduced by the opening of the superconducting energy gap.

\end{abstract}


\maketitle

\section{\label{sec:intro}INTRODUCTION}

The superconducting transition always breaks gauge symmetry, which is the only broken symmetry in ``conventional'' superconductors. Unconventional superconductivity is characterized by additional broken symmetries, including time-reversal symmetry (TRS)~\cite{Sigrist91,Mineev99}. Broken TRS in superconductors, which is quite rare, is especially interesting, because it implies not just unconventional pairing, but also the existence of twofold or higher degeneracy of the superconducting order parameter. The detection of a spontaneous but very small internal field~$B_s$ below the superconducting transition temperature $T_c$ in a number of superconductors~\cite{Heffner90,Luke93,Luke98,Hillier09,Biswas13,Singh14,Hillier12,ATKS03,Shu11,Maisuradze10} is strong experimental evidence for broken TRS\@.

Zero-field muon spin relaxation (ZF-$\mu$SR) is especially sensitive to small changes in internal fields and can often measure fields of 0.01~mT, corresponding to $10^{-2}\text{--}10^{-3}\mu_{\rm B}$ if produced by dipolar coupling to a lattice of local moments. This makes ZF-$\mu$SR an extremely powerful technique for discovering and characterizing TRS breaking in exotic superconductors. Spontaneous fields~$B_s$ have been observed by ZF-$\mu$SR in the heavy-fermion superconductors~(U,Th)Be$_{13}$~\cite{Heffner90} and UPt$_3$~\cite{Luke93} (although not without controversy~\cite{Dalmas95,Higemoto00}; see also \cite{Schemm14}), the candidate chiral $p$-wave superconductor Sr$_2$RuO$_4$~\cite{Luke98}, the non-centrosymmetric superconductors LaNiC$_2$~\cite{Hillier09}, SrPtAs~\cite{Biswas13}, and Re$_6$Zr~\cite{Singh14}, the centrosymmetric superconductor LaNiGa$_2$~\cite{Hillier12}, and the filled skutterudite superconductors~(Pr,La)(Os,Ru)$_4$Sb$_{12}$~\cite{ATKS03,Shu11} and PrPt$_4$Ge$_{12}$~\cite{Maisuradze10}.

The ratios of the superconducting gaps to $k_BT_c$ in PrOs$_4$Sb$_{12}$ ($T_c = 1.8$~K)~\cite{BFHZ02} and PrPt$_4$Ge$_{12}$ ($T_c = 7.9$~K)~\cite{Gumeniuk08} are similar, but their crystalline-electric-field (CEF) level splitting schemes are quite different. Both have the same nonmagnetic singlet $\Gamma_1$ ground state, but in PrOs$_4$Sb$_{12}$ the first excited triplet $\Gamma_4^{(2)}$ CEF-split state (splitting $\sim$8~K) strongly hybridizes with the ground state and conduction electrons~\cite{Maple07}, generating a heavy-fermion state, whereas in PrPt$_4$Ge$_{12}$ the first excited CEF state is a different triplet ($\Gamma_4^{(1)}$ in $T_h$ notation), and the splitting is much larger (120-130~K)~\cite{Gumeniuk08, Toda08}. Heavy-fermion behavior is not observed in thermodynamic data for PrPt$_4$Ge$_{12}$~\cite{Gumeniuk08}.

ZF-$\mu$SR measurements in both PrOs$_4$Sb$_{12}$ and PrPt$_4$Sb$_{12}$ are consistent with a superconducting state that breaks TRS~\cite{ATKS03,Maisuradze10}, although to date neither the detailed symmetry of the pairing nor its irreducible representation have been well determined. ZF-$\mu$SR experiments in the Pr(Os,Ru)$_4$Sb$_{12}$ and (Pr,La)Os$_4$Sb$_{12}$ alloy series~\cite{Adroja05,Shu11} suggest that broken TRS is suppressed for Ru concentration $\geqslant 0.6$ but persists up to La concentration $\approx 1$, and support a crystal-field excitonic Cooper pairing mechanism for TRS-breaking superconductivity~\cite{Shu11}.

A detailed study of the evolution of the superconducting and normal state properties of (Pr,Ce)Pt$_4$Ge$_{12}$ raises interesting questions about broken TRS in PrPt$_4$Ge$_{12}$~\cite{Huang14}. Superconductivity is suppressed with increasing Ce with positive curvature up to $x = 0.4$, above which no evidence for superconductivity was observed down to 1.1~K\@. From specific heat measurements it was shown that the electron correlations are enhanced with increasing Ce concentration. The $C(T)/T$ data in the superconducting state are best described by a $T^3$ dependence for $x=0$~\cite{Maisuradze09,Huang14} and an $e^{-\Delta/T}$ dependence for $x \gtrsim 0.05$~\cite{Huang14}, indicating a crossover from a nodal to nodeless superconducting energy gap or the suppression from multiple to single BCS type superconducting energy bands with increasing Ce concentration. This crossover motivated the current investigation on the evolution of broken TRS in PrPt$_4$Ge$_{12}$ with Ce substitution.

In this Article we report the results of ZF-$\mu$SR experiments in Pr$_{1-x}$Ce$_x$Pt$_4$Ge$_{12}$, which were undertaken to study the evolution of the spontaneous local field $B_s$ below $T_c$ with Ce doping. A linear decrease of $B_s$ with Ce concentration is observed up to $x = 0.2$. Our results suggest that $B_s$ is suppressed to zero at $x \approx 0.4$, which is near the critical concentration for suppression of $T_c$ to zero. This resembles the behavior of $B_s$ in (Pr,La)Os$_4$Sb$_{12}$, where broken TRS is associated directly with the Pr concentration, more than in Pr(Os,Ru)$_4$Sb$_{12}$, where the Pr concentration is unchanged~\cite{Shu11}.

\section{\label{sec:expt} EXPERIMENTAL}

Powder samples of polycrystalline Pr$_{1-x}$Ce$_{x}$Pt$_4$Ge$_{12}$ with $x = 0$, 0.07, 0.1, and 0.2 were synthesized as described in Ref.~\cite{Huang14}. Rietveld refinements were conducted on powder XRD patterns for each sample. The body centered cubic structure with space group $Im\bar{3}$ was observed, consistent with that reported in the literature~\cite{Gumeniuk10,Chandra12}. ZF-$\mu$SR experiments were carried out on at the ISIS Neutron and Muon Facility, Rutherford Appleton Laboratory, Chilton, U.K.

\begin{figure}[ht]
 \begin{center}
 \includegraphics[width=3.25in]{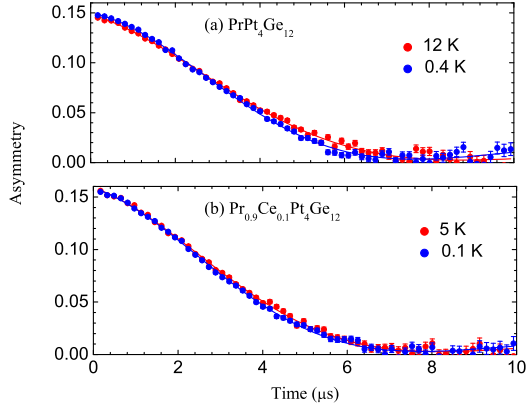}
 \caption{(Color online) Time evolution of $\mu^+$ decay positron asymmetry, proportional to the $\mu^+$ spin polarization $P_{\mu}(t)$, above and below the superconducting transition in (a) PrPt$_4$Ge$_{12}$ and (b) Pr$_{0.9}$Ce$_{0.1}$Pt$_4$Ge$_{12}$. A constant signal from muons that miss the sample and stop in the silver sample holder has been subtracted from the data.}
 \label{fig:asy}
 \end{center}
\end{figure}

Figure~\ref{fig:asy} shows the time evolution of the decay positron count rate asymmetry, proportional to the positive-muon ($\mu^{+}$) spin polarization $P_{\mu}(t)$~\cite{Brewer94}, in Pr$_{1-x}$Ce$_x$Pt$_4$Ge$_{12}$, $x = 0$ and 0.1, at temperatures above and below $T_c$.
A constant background signal, which originates from muons that miss the sample and stop in the silver sample holder, has been subtracted from the data. As previously reported by Maisuradze {\em et al.}~\cite{Maisuradze10}, in the end compound PrPt$_{4}$Ge$_{12}$ there is a small but resolved increase in relaxation rate in the superconducting state. Similar but smaller increases are observed in the Ce-doped alloys.

We initially fit our data using an exponentially damped version of the ``golden formula" of Kubo~\cite{Kubo81} or ``Voigtian''~\cite{Maisuradze10} function:
\begin{equation}
 \label{eq:dGGKT}
 P_{\mu}(t)=\exp(-\Lambda t)G_z ^{\rm K-T}(\Delta,\lambda,t),
 \end{equation}
where
\begin{equation}
 \label{eq:GKT}
 G_{z}^{\rm K-T}(\Delta,\lambda, t)=\frac{1}{3}+\frac{2}{3}(1-\Delta^2t^2-\lambda t)\exp(-\frac{1}{2}\Delta^2t^2-\lambda t).
\end{equation}
Equation~(\ref{eq:GKT}) describes a convolution of Gaussian and Lorentzian distributions of randomly-oriented static (or quasistatic) local fields at $\mu^+$ sites with distribution widths~$\delta B_G$ (the rms width) and $\delta B_L$, respectively; the relaxation rates~$\Delta$ and $\lambda$ are defined by $\Delta =\gamma_\mu \delta B_G$ and $\lambda= \gamma_\mu \delta B\lambda$, where $\gamma_\mu = 2\pi\times135.53$~MHz/T is the $\mu^+$ gyromagnetic ratio. In Eq.~(\ref{eq:dGGKT}) the exponential damping with rate~$\Lambda$ models dynamic relaxation by a fluctuating additional contribution to the local field. In contrast to the results of Ref.~\cite{Maisuradze10}, we find extremely small values of $\lambda$, and furthermore the increase of $\Delta$ below $T_c$ is the same as when $\lambda$ is set fixed to zero. Thus the simpler damped Gaussian Kubo-Toyabe function~\cite{Hayano79}
\begin{equation}
 \label{eq:dGKT}
 P_{\mu}(t)=\exp(-\Lambda t)G_z ^{\rm K-T}(\Delta,t),
 \end{equation}
where
\begin{equation}
 \label{eq:KT}
 G_{z}^{\rm K-T}(\Delta, t)=\frac{1}{3}+\frac{2}{3}(1-\Delta^2t^2)\exp(-\frac{1}{2}\Delta^2t^2)
\end{equation}
(i.e., the assumption that the $\mu^+$ local field distribution is purely Gaussian with rms width $\Delta/\gamma_\mu$) describes the data adequately. Equation~(\ref{eq:dGKT}) was used previously to fit ZF-$\mu$SR data from Pr(Os,Ru)$_4$Sb$_{12}$ and (Pr,La)Os$_4$Sb$_{12}$ \cite{SMAT07}. We also fit the present data using the so-called ``dynamic'' K-T function~\cite{Hayano79} that models local-field fluctuations with full reorientation (fits not shown), but the fits are poorer than those to Eq.~(\ref{eq:dGKT}).

\section{\label{sec:results} RESULTS}

\begin{figure}[ht]
 \begin{center}
 \includegraphics[width=3.25in]{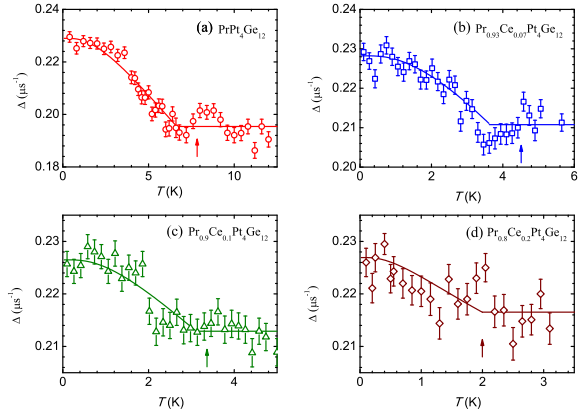}
 \caption{(Color online) Temperature dependence of the ZF Kubo-Toyabe static relaxation rate $\Delta$ in (a)~PrPt$_4$Ge$_{12}$, (b)~Pr$_{0.93}$Ce$_{0.07}$Pt$_4$Ge$_{12}$, (c)~Pr$_{0.9}$Ce$_{0.1}$Pt$_{4}$Ge$_{12}$, and (d)~Pr$_{0.8}$Ce$_{0.2}$Pt$_{4}$Ge$_{12}$. Curves: fits of Eq.~(\ref{eq:Delta}) assuming the temperature dependence of the BCS order parameter for $\Delta_e(T)$. Arrows: superconducting transition temperature $T_c^{Cp}$ from specific heat measurements~\protect\cite{Huang14}.}
 \label{fig:rlx}
 \end{center}
\end{figure}

Figure~\ref{fig:rlx} shows the temperature dependence of $\Delta$ in Pr$_{1-x}$Ce$_{x}$Pt$_4$Ge$_{12}$, $x=0, 0.07$, 0.1, and 0.2.
An increase of $\Delta$ below the superconducting transition temperature~$T_c^{C_p}$ determined from the specific heat~\cite{Huang14} is observed in all alloys, indicating the onset of a spontaneous field $B_s$ in the superconducting state. The size of this increase decreases with increasing Ce concentration. In the end compound~PrPt$_4$Ge$_{12}$ the increase starts around 6.7~K, as in Ref.~\cite{Maisuradze10}, but the size of the increase shown in Fig.~\ref{fig:rlx}(a) is greater than that reported by these authors.

The nuclear dipolar and electronic contributions to $\Delta$ in the superconducting state are uncorrelated and added in quadrature~\cite{ATKS03}:
\begin{equation} \label{eq:Delta}
 \Delta(T) = \left\{ \begin{array}{cc}
 \Delta_n \,, & T > T_c \,, \\
 \left[\Delta_n^2+\Delta_e^2(T)\right]^{1/2}, & T < T_c \,
 \end{array}
\right.
\end{equation}
where $\Delta_n/\gamma_\mu$ is the temperature-independent rms nuclear dipolar field distribution width and $\Delta_e(T)/\gamma_\mu$ is the width of the spontaneous field distribution from broken TRS~\cite{ATKS03} that we associate with $B_s$. Equation~(\ref{eq:Delta}) was fitted to the data of Fig.~\ref{fig:rlx} assuming $\Delta_e$ has the temperature dependence of the BCS order parameter, for which we use the approximate empirical expression
\begin{equation}
 \label{eq:BCS}
 \Delta_e(T)=\Delta_e(0) \tanh\left[b\sqrt{\frac{T_c}{T}-1}\right];
\end{equation}
here $b$ is a dimensionless coefficient ($b = 1.74$ for an isotropic BCS superconductor in the weak-coupling limit)~\cite{Gross86}. The amplitude $\Delta_e(0)$ of $\Delta_e(T)$, $b$, $T_c$, and $\Delta_n$ were varied for best fit. (For $x = 0.2$, $\Delta_e(0)$ becomes too small to determine $T_c$ from the fit, and $T_c$ was fixed at $T_c^{C_p}$.)

The values of the parameters from the fits are shown in Table~\ref{tabel}\@.
\begin{table}[ht]
\caption{Parameters from fits of Eqs.~(\protect\ref{eq:Delta}) and (\protect\ref{eq:BCS}) to the data of Fig.~\ref{fig:rlx}. $T_c^{C_p}$: superconducting transition temperature from specific heat measurements~~\protect\cite{Huang14}.}
\label{tabel}
\begin{ruledtabular}
\begin{tabular}{lcccc}
Ce concentration $x$ & 0 & 0.07 & 0.1 & 0.2 \\
\hline
 $\Delta_{n}\ (\mu\text{s}^{-1})$ & 0.195(4) & 0.211(1)&0.213(1) & 0.216(4) \\
 $\Delta_{e}(0)\ (\mu\text{s}^{-1})$ & 0.120(3) & 0.087(3)&0.077(3)& 0.068(4) \\
 $\Delta_{e}(0)/\gamma_\mu$ (mT) & 0.141(4) & 0.102(4)&0.090(4)& 0.080(5) \\
 $b$ &1.2(1) & 1.3(2) &1.3(2) & 1.1(4) \\
 $T_c$ (K) & 6.7(3) & 3.6(2) &3.1(1) & 2.0 \\
$T_c^{C_p}$ (K) & 7.9 & 4.5 & 3.4 & 2.0 \\
 \end{tabular}
\end{ruledtabular}
\end{table}
To within error, $b$ is independent of Ce concentration $x$ and smaller than the isotropic BCS value. As shown in Fig.~\ref{fig:rlx}, the rise of $\Delta$ begins somewhat below $T_c^{Cp}$, so that for $x=0$, 0.07, and 0.1 $T_c$ is smaller than $T_c^{C_p}$. There is no indication for a phase transition below $T_c^{C_p}$ from bulk measurements~\cite{Huang14}.

 The magnitude of $B_s$ is difficult to estimate theoretically~\cite{ATKS03}. The uniform spin and orbital fields expected for non-unitary pairing~\cite{OhMa93} are ${\lesssim}10^{-3}$~mT for PrPt$_4$Ge$_{12}$, and therefore negligible compared to $\Delta_e(0)/\gamma_\mu$ (Table~\ref{tabel}). Fields produced by inhomogeneity of the superconducting order parameter due to lattice defects, impurities, etc.~\cite{Choi89,Mineev89} depend strongly on the nature and density of such defects~\cite{Sigrist91} (which might explain the difference between our results and those of Ref.~\cite{Maisuradze10}). Very rough estimates from the results of \cite{Choi89, Mineev89} for the field at an impurity site (which is of course not the muon site) are of the order of 0.01~mT, an order of magnitude smaller than our values of $\Delta_e(0)/\gamma_\mu$.

A striking difference between (PrOs$_4$Sb$_{12}$)- and (PrPt$_4$Ge$_{12}$)-based materials is the fact that in the former alloy series the observed quasistatic relaxation in the normal state is accounted for by $^{121}$Sb and $^{123}$Sb nuclear dipolar fields~\cite{SMAT07}, whereas in (Pr,Ce)Pt$_4$Ge$_{12}$ latter none of the bare (i.e., unenhanced) nuclear magnetic moments are large enough to do this. The largest contribution is from $^{141}$Pr nuclei, for which a simple lattice-sum second moment calculation~\cite{Hayano79} yields $\Delta_n(\text{bare}) \approx 0.04~\mu\text{s}^{-1}$ assuming the $\mu^+$ site reported in Ref.~\cite{ATKS03}.

Comparison with measured values of $\Delta_n$ (Table~\ref{tabel}) shows that $^{141}$Pr hyperfine enhancement by about 5 is required. The enhancement factor~$K$ is given by $K = a_\mathrm{hf}\chi_\mathrm{mol}$~\cite{Bleaney73,SMAT07}, where $a_\mathrm{hf} = 187.7$~mole/emu is the Pr atomic hyperfine coupling constant. For PrPt$_4$Ge$_{12}$ $\chi_\mathrm{mol} = 23 \times 10^{-3}$~emu/mole-Pr at low temperatures~\cite{Maisuradze10} so that $K \approx 4.3$. This is close to the required value, although uncertainties in the anisotropy of the hyperfine enhancement and the $\mu^+$ site prevent a detailed comparison. We conclude that dipolar fields from hyperfine-enhanced $^{141}$Pr nuclei are responsible for the quasistatic component of the $\mu^+$ spin relaxation in the normal state.

A small dip in $\Delta(T)$ is observed just below $T_c^{C_p}$ for $x = 0$, 0.07, and 0.1, as previously reported for PrPt$_4$Ge$_{12}$~\cite{Maisuradze10}. These authors speculated that this might be due to diluted magnetic centers separated by distances of order of the magnetic penetration depth~$\lambda_L = 114(4)$~nm~\cite{Maisuradze09}, so that $\Delta$ is reduced due to screening of the impurity magnetic field. Such impurities were not observed, however, based on the absence of a low-temperature upturn in the magnetic susceptibilities down to $\sim$7~K~\cite{Maisuradze10}\@. It should also be noted that such screening requires an impurity concentration~$c_\mathrm{imp} \lesssim 1/\lambda_L^3 \approx 6.8\times10^{14}~\text{cm}^{-3}$. This concentration is extremely small (${\sim}2\times10^{-7}$/Pr ion). The dipolar field at a $\mu^+$ site of the order of this distance from an impurity is $\sim\mu_\mathrm{imp}/\lambda_L^3 \approx 6\times10^{-10}~\text{T}/\mu_B$. which is negligible compared to observed values of $\Delta/\gamma_\mu \sim 0.1$~mT from Table~\ref{tabel}\@. We conclude that magnetic impurities cannot account for the dip, and its origin remains unknown.

\begin{figure}[ht]
 \begin{center}
 \includegraphics[width=3.25in]{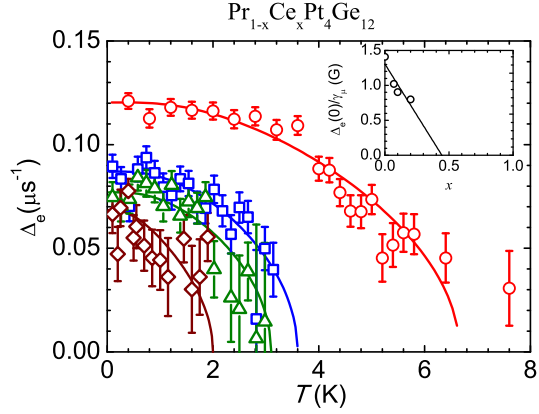}
 \caption{(Color online) Points: temperature dependence of the relaxation rate $\Delta_{e}$ in (Pr,Ce)Pt$_4$Ge$_{12}$ (circles: $x=0$, squares: $x=0.07$, triangles: $x=0.1$, and diamonds: $x = 0.2$). Curves: fits of Eq.~(\ref{eq:BCS}) to the data. Inset: dependence of the rms width width $\Delta_e(0)/\gamma_{\mu}$ of the $T=0$ spontaneous field distribution on Ce concentration $x$ in Pr$_{1-x}$Ce$_{x}$Pt$_{4}$Ge$_{12}$. Solid line: linear fit.}
 \label{fig:rlx_e}
 \end{center}
\end{figure}

Figure~\ref{fig:rlx_e} shows the temperature dependence of the relaxation rate $\Delta_{e}$ in (Pr,Ce)Pt$_4$Ge$_{12}$ obtained by solving Eq.~(\ref{eq:Delta}).
The dependence of $\Delta_e(0)/\gamma_{\mu}$ on $x$ is shown in the inset. A linear fit suggests that TRS is suppressed for $x \approx 0.4$. This is also the critical Ce concentration for which superconductivity is suppressed~\cite{Huang14}. The consequences of this are discussed briefly in Sec.~\ref{sec:concl}.

\begin{figure}[ht]
 \begin{center}
 \includegraphics[width=3.25in]{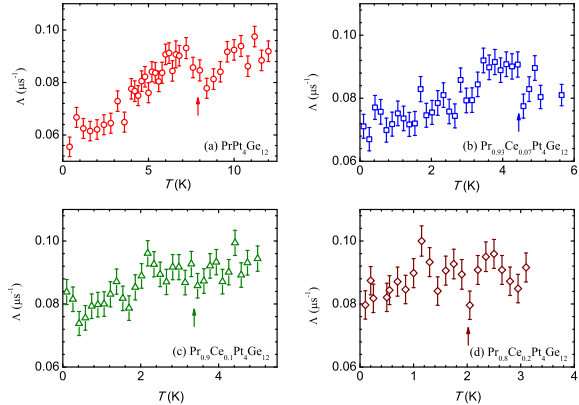}
 \caption{(Color online) Temperature dependence of the ZF exponential damping rate $\Lambda$ in Pr$_{1-x}$Ce$_{x}$Pt$_4$Ge$_{12}$. (a)~$x=0$. (b)~$x=0.07$. (c)~$x=0.1$. (d)~$x = 0.2$. Arrows: $T_c^{C_p}$ from specific heat measurements~\protect\cite{Huang14}.}
 \label{fig:Lam}
 \end{center}
\end{figure}

Figure~\ref{fig:Lam} shows the temperature dependence of the dynamic rate~$\Lambda$ in (Pr,Ce)Os$_4$Ge$_{12}$.
There is some indication of a weak temperature dependence of $\Lambda$ in the normal state, although the uncertainty is large. Below $T_c$ $\Lambda$ decreases with decreasing temperature, most strongly in the end compound PrPt$_4$Ge$_{12}$. We note that $\Delta$ and $\Lambda$ are anticorrelated in fitting the data to Eq.~(\ref{eq:dGKT}), so that the increased $\Delta$ below $T_c$ could result from a decrease in $\Lambda$ (or vice versa). There seems to be some anticorrelation in the neighborhood of $T_c$, particularly for $x = 0$ (cf.\ Figs.~\ref{fig:rlx} and \ref{fig:Lam}). However, the asymmetry data exhibit a qualitative increase in relaxation rate below $T_c$ (Fig.~\ref{fig:asy}), and fits to the data with $\Lambda$ held fixed (not shown) also yield increases in $\Delta$.

In both Pr(Os,Ru)$_4$Sb$_{12}$ and (Pr,La)Os$_4$Sb$_{12}$ the exponential damping rate~$\Lambda$ was found to increase slightly with decreasing temperature with no evidence for an anomaly at $T_c$~\cite{SMAT07}. The trend is different in (Pr,Ce)Pt$_4$Ge$_{12}$, where $\Lambda$ decreases significantly below $T_c$ (Fig.~\ref{fig:Lam}), at least for the lower Ce concentrations (the decrease is smaller for $x = 0.1$ and 0.2, making it harder to detect the anomaly).

Hyperfine-enhanced dipolar fields from $^{141}$Pr nuclear spin fluctuations were suggested as the origin of the $\mu^+$ dynamic relaxation in Pr(Os,Ru)$_4$Sb$_{12}$ and (Pr,La)Os$_4$Sb$_{12}$~\cite{SMAT07}. In those materials, nuclear spin dynamics appear to be driven by hyperfine-enhanced nuclear spin-spin interactions that are not strongly affected by superconductivity. Thus the temperature dependence of $\Lambda$ in (Pr,Ce)Pt$_4$Ge$_{12}$ cannot be accounted for by this mechanism, and in addition the hyperfine enhancement is reduced by two orders of magnitude by the increased Pr$^{3+}$ CEF splitting in (Pr,Ce)Pt$_4$Ge$_{12}$. Another explanation for the dynamic relaxation and its temperature dependence is necessary.

The decrease of $\Lambda$ below $T_c$ might be due to opening of the superconducting gap. The $^{73}$Ge nuclear spin-lattice relaxation rate~$1/^{73}T_1(T)$, measured using zero-field NQR~\cite{Kanetake10}, shows this effect clearly. It is striking that in PrPt$_4$Ge$_{12}$ both $\Lambda(T)$ [Fig.~\ref{fig:Lam}(a)] and $1/^{73}T_1(T)$~\cite{Kanetake10} exhibit a maximum just below $T_c$ that resembles the Hebel-Slichter ``coherence'' peak expected in a superconductor with an isotropic gap~\cite{Hebel59}. At lower temperatures, however, $1/^{73}T_1(T)$ decreases exponentially~\cite{Kanetake10}, whereas $\Lambda$ remains nonzero down to 25~mK (Fig.~\ref{fig:Lam}). Furthermore, conduction-electron Korringa relaxation is rarely visible in $\mu$SR, since the $\mu^+$--conduction-electron hyperfine interaction is weak and the resulting relaxation times are usually much longer than the $\mu^+$ lifetime.

Alternatively, dynamic $\mu^+$ spin relaxation might arise from fluctuations of $^{141}$Pr nuclear dipolar fields due to Korringa relaxation of the Pr nuclei, which is reduced by the opening of the superconducting gap. In Pr(Os,Ru)$_4$Sb$_{12}$ and (Pr,La)Os$_4$Sb$_{12}$ the dynamic muon spin relaxation is provided by fluctuating $^{141}$Pr dipolar fields, with quasistatic local fields supplied by Sb nuclei~\cite{Adroja05,SMAT07}. In contrast, in (Pr,Ce)Pt$_4$Ge$_{12}$ the only appreciable nuclear dipolar fields are from $^{141}$Pr nuclei. If their fluctuations are rapid the quasistatic field is averaged to zero, leaving a single-exponential $\mu^+$ spin relaxation function contrary to experiment (Fig.~\ref{fig:asy}). If on the other hand the $^{141}$Pr fluctuations are slow (``adiabatic''), then the $\mu^+$ and $^{141}$Pr fluctuation rates are nearly the same~\cite{Hayano79}.

In this scenario the opening of the superconducting gap reduces the $^{141}$Pr Korringa relaxation rate, which is then mirrored by $\Lambda$. This is consistent with the data. As noted above, however, the dynamic K-T relaxation function appropriate to this ``single-field-source'' picture does not fit the data as well as the damped static K-T function of Eq.~(\ref{eq:dGKT}) that assumes two $\mu^+$ local field sources: one quasistatic (the hyperfine-enhanced $^{141}$Pr dipolar field), and the other fluctuating (the putative conduction-electron hyperfine interaction). Thus it is difficult to decide between these two possibilities, and the mechanism for dynamic $\mu^+$ spin relaxation in Pr$_{1-x}$Ce$_x$Pt$_4$Ge$_{12}$ is not yet fully understood.

\section{\label{sec:concl} CONCLUSIONS}

ZF-$\mu$SR measurements on Pr$_{1-x}$Ce$_x$Pt$_4$Ge$_{12}$ show that broken TRS in PrPt$_4$Ge$_{12}$ is suppressed by Ce doping. The spontaneous magnetic field that signals broken TRS decreases linearly with $x$ and $\to 0$ at $x \approx 0.4$, which is near the critical concentration for which the superconducting transition temperature is suppressed to zero~\cite{Huang14}. In this respect the results resemble those from (Pr,La)Os$_4$Sb$_{12}$, for which the Pr sublattice is also diluted, except that in the latter alloy series the end compound LaOs$_4$Sb$_{12}$ is also superconducting and there is a crossover between superconducting ground states with broken and non-broken TRS~\cite{Shu11}.

In (Pr,Ce)Pt$_4$Ge$_{12}$ both broken TRS and superconductivity itself are suppressed above a critical Ce concentration $x_\mathrm{cr} \approx 0.4$. This differs from the situation in (Pr,La)Os$_4$Sb$_{12}$, where the proportionality of $\Delta_e(0)$ to the Pr concentration indicates that Pr-Pr interactions are responsible for the broken TRS, and in Pr(Os,Ru)$_4$Sb$_{12}$, where the data suggest that the increase of the CEF excitation energy with Ru concentration is driving the restoration of TRS~\cite{Shu11}. The reduction of $T_c$ in (Pr,Ce)Pt$_4$Ge$_{12}$ appears to be driven by a pair-breaking effect of the Ce doping on the remaining Pr ions, in addition to the weakening of Pr-Pr coupling by dilution.

The reduction of the dynamic $\mu^+$ spin relaxation rate~$\Lambda$ below $T_c$ (Fig.~\ref{fig:Lam}) seems to reflect the opening of the superconducting gap. This suggests that conduction electrons contribute to $\Lambda$ via the Korringa mechanism. Observation of Korringa relaxation in $\mu^+$SR is unusual, and details of the required $\mu^+$--conduction-band interaction remain unclear; more work is required to elucidate this behavior.

\begin{acknowledgments}
We are grateful to the STFC for beam time at the ISIS facility, and to the ISIS Cryogenics Group for invaluable help during the experiments. This research is supported by the National Natural Science Foundation of China (11204041 and 11474060), Natural Science Foundation of Shanghai, China (12ZR1401200), STCSM of China (No.15XD1500200), the U.S. Department of Energy (DOE) under Research Grant No.~DE-FG02-04ER46105 (sample synthesis at UCSD) and the U.S. National Science Foundation under Grant No.~1206553 (sample characterization at UCSD).

\end{acknowledgments}


\end{document}